\documentclass[a4paper]{article}

\usepackage{ISCSLP_v2}
\usepackage{color,xcolor}
\newcommand{\tabincell}[2]{\begin{tabular}{@{}#1@{}}#2\end{tabular}}

\title{Hybrid CTC-Attention based End-to-End Speech Recognition \\ using Subword Units}

\name{Zhangyu Xiao$^1$, Zhijian Ou$^1$, Wei Chu$^2$, Hui Lin$^2$\thanks{This  work  is  supported  by  NSFC  grant  61473168  and a Liulishuo grant. Correspondence to: Z. Ou (ozj@tsinghua.edu.cn).}}
\address{
  $^1$Department of Electronic Engineering, Tsinghua University, Beijing 100084, China\\
  $^2$Shanghai Liulishuo Information Technology Co., Ltd}
\email{xiaozy13@mails.tsinghua.edu.cn, ozj@tsinghua.edu.cn, \{wei.chu,hui.lin\}@liulishuo.com}

\begin{document}

\maketitle
\begin{abstract}
In this paper, we present an end-to-end automatic speech recognition system, which successfully employs subword units in a hybrid CTC-Attention based system.
The subword units are obtained by the byte-pair encoding (BPE) compression algorithm.
Compared to using words as modeling units, using characters or subword units does not suffer from the out-of-vocabulary (OOV) problem.
Furthermore, using subword units further offers a capability in modeling longer context than using characters.
We evaluate different systems over the LibriSpeech 1000h dataset.
The subword-based hybrid CTC-Attention system obtains 6.8\% word error rate (WER) on the test\_clean subset without any dictionary or external language model.
This represents a significant improvement (a 12.8\% WER relative reduction) over the character-based hybrid CTC-Attention system.

\end{abstract}

\section{Introduction}

Traditional large vocabulary continuous speech recognition (LVCSR) systems consist of a complex pipeline of multiple modules, such as a GMM/DNN-HMM based acoustic model, a pronunciation lexicon and an external word-level language model \cite{rabiner1989tutorial, hinton2012deep}. Building such a complex LVCSR system remains a complicated task, which requires expertise-intensive knowledge.

In this paper, we are interested in building end-to-end speech recognition systems, which have shown promising results on LVCSR tasks. An end-to-end system generally denotes a simplified pipeline, which is usually based on neural network architectures and can be trained from scratch.
Generally, there are two main approaches for designing end-to-end LVCSR systems: Connectionist Temporal Classification (CTC) \cite{graves2006connectionist} and RNN encoder-decoder with attention mechanism \cite{bahdanau2016end}.

However, these end-to-end LVCSR systems still have several drawbacks. On the one hand, CTC makes a strong independent assumption between labels, thus cannot perform well without a strong language model. Attention based methods solve this problem by training a decoder which emits labels depending on previous ones. On the other hand, attention based system are hard to train due to its excessively flexible attention alignments which might be unreasonable. In speech recognition tasks, the alignment between input features and output symbols is usually monotonic. CTC uses the so-called latent path to represent this alignment. Recent work that combines CTC and attention loss in the end-to-end system \cite{kim2017joint} achieves lower WERs than using either approach individually.
This is our first observation to improve end-to-end speech recognition system.

Our second observation is concerned with different choices of the modeling units.
End-to-end systems directly map acoustic features to label sequences, which are composed of symbols like phonemes \cite{graves2014towards, miao2015eesen}, characters \cite{bahdanau2016end, kim2017joint, collobert2016wav2letter, chan2016listen, maas2015lexicon, amodei2016ds2}, subwords \cite{rao2017exploring, zenkel2017subword} and words \cite{soltau2016neural}.
Phoneme based approaches need a carefully designed pronunciation lexicon to map words to phoneme sequences.
Both phoneme based models \cite{graves2014towards, miao2015eesen} and word based models \cite{soltau2016neural} need a predefined dictionary, thus cannot handle the out-of-vocabulary (OOV) problem.
In contrast, characters, or graphemes, are advantageous for end-to-end systems, since all text can be easily segmented into character sequence, and thus naturally enable open vocabulary end-to-end speech recognition.
However, using characters would increase the burden in learning longer context dependency.
Using subword units would potentially overcome these drawbacks, and still keep the advantage for open vocabulary end-to-end recognition.
Recently, the subword based model has shown impressive results in neural machine translation (NMT) \cite{sennrich2015neural, wu2016google} because of its ability to deal with infrequent words, like compounds, cognates as well as loanwords. For end-to-end speech recognition, there are also successful applications with subword units \cite{rao2017exploring, zenkel2017subword}.

In \cite{zenkel2017subword}, both subword units and cross-word units are generated with the byte-pair encoding (BPE) \cite{sennrich2015neural} method, and the neural network is trained based on the CTC loss using a subword and cross-word based language model.
Cross-word units are taken into the unit set in order to model liaisons in oral English conversations, such as speaking ``gonna'' instead of ``going to''. However, the CTC model employed in \cite{zenkel2017subword} performs poorly without an external language model.
Using an external language model would require a predefined dictionary.
In \cite{rao2017exploring}, the authors adopt the word-piece model (WPM) \cite{schuster2012japanese} to produce their subword units and employs the RNN-Transduer (RNN-T) neural architecture.
It is shown in \cite{rao2017exploring} that RNN-T with WPM significantly outperforms the character based RNN-T.
The encoder network is pre-trained with a CTC loss while the decoder network is initialized with a pre-trained LSTM language model. Without careful tuning of pre-training, it is nearly impossible to train an effective RNN-T.
Note that both BPE and WPM based subword models use a fixed decomposition of words \cite{rao2017exploring, zenkel2017subword}. Learning a variable decomposition of the target sequence is studied in \cite{liu2017gram, chan2016latent}.
In \cite{liu2017gram}, the authors propose a method called GRAM-CTC to jointly learn the alignment between the input and output sequence as well as a better decomposition for the target sequence. In \cite{chan2016latent}, the authors add an addtional objective of learning a dynamic decomposition to the training loss function. Both variable decomposition approaches greatly increase the difficulty of training, and thus require more careful fine-tuning and higher computational cost.

From the above two observations, we present an end-to-end LVCSR system in this paper, which successfully employs subword units in a hybrid CTC-Attention neural architecture.
The combination of subword modeling and hybrid CTC-Attention has not been explored, to the best of our knowledge, which would contribute to a stronger end-to-end system.
We use the BPE method to construct the subword units, and evaluate different systems over the LibriSpeech 1000h dataset \cite{panayotov2015librispeech}. The subword-based hybrid CTC-Attention system obtains 6.8\% word error rate (WER) on the test\_clean subset without any dictionary or external language model.
This represents a significant improvement (a 12.8\% WER relative reduction) compared to the character-based hybrid CTC-Attention system.

\section{Hybrid CTC-Attention Architecture}

\begin{figure}[t]
	\centering
	\includegraphics[width=\linewidth]{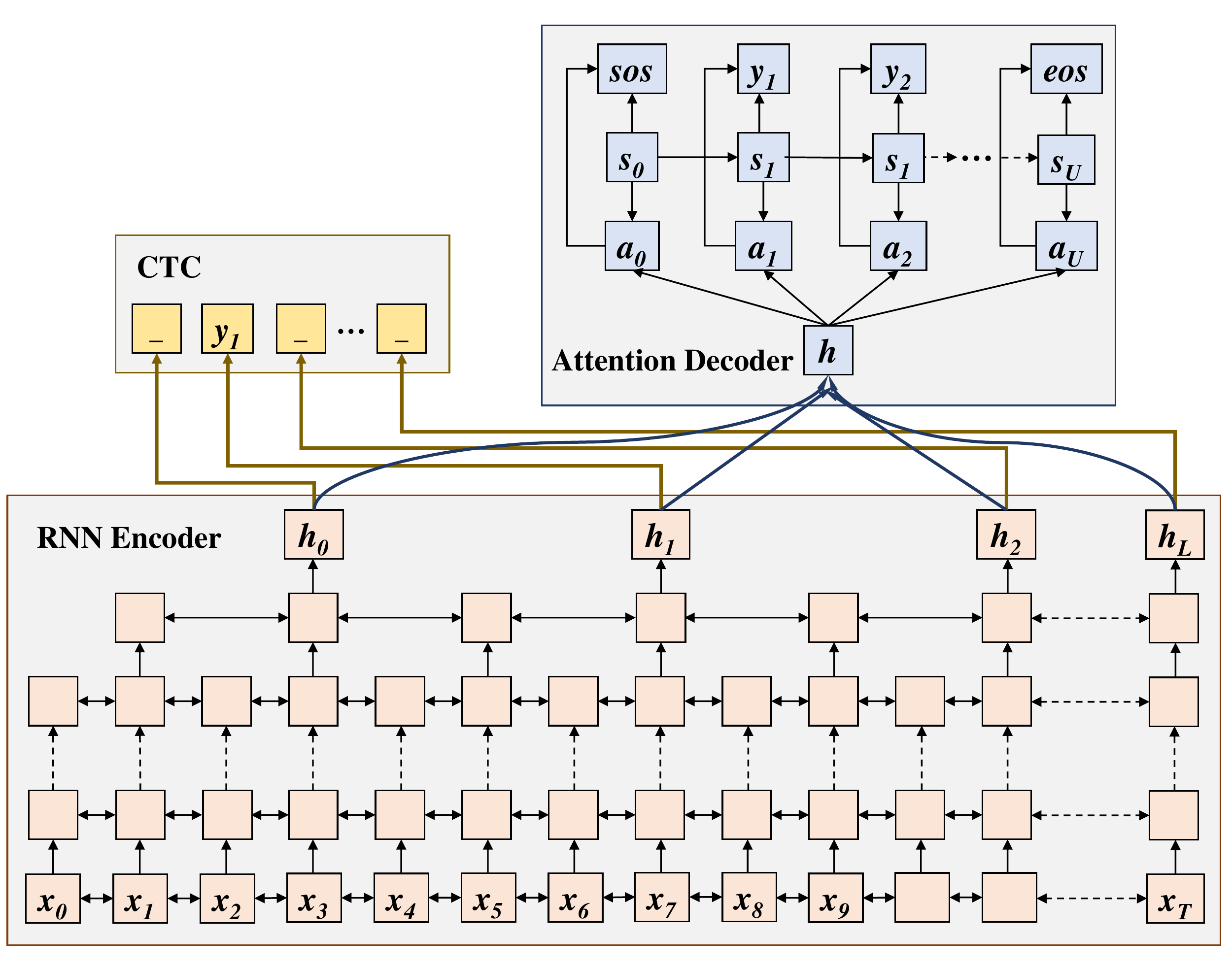}
	\caption{The hybrid CTC-Attention model consists of three modules: RNN Encoder, CTC Loss and Attention Decoder. RNN Encoder is implemented by stacking multiple BLSTM layers, in which the top two layers subsample the hidden states from layers below by a factor of 2. CTC and Attention Decoder share the same output deep features from RNN Encoder and compute objective functions simultaneously.}
	\label{fig:speech_production}
\end{figure}

In this section, we introduce a hybrid CTC-Attention architecture \cite{kim2017joint} for end-to-end speech recognition. The overall architecture can be found in Figure \ref{fig:speech_production}. Given an input acoustic feature sequence \(x=(x_1,x_2,\ldots,x_T)\) and an output symbol sequence \(y=(y_1,y_2,\ldots,y_U)\), the hybrid CTC-Attention based framework models the transcription between \(x\) and \(y\). Note that \(y_u\in\{1,\ldots,K\}\), $K$ is the number of different label units. In end-to-end speech recognition systems, the length of output label sequence is usually shorter than the length of feature sequence (i.e., \(U<T)\).

The hybrid CTC-Attention architecture uses a shared RNN encoder to produce a high-level representation \(h=(h_1,h_2,\ldots,h_L)\)  for the input sequence \(x\), where \(L\) is the downsampled frame index:
\begin{displaymath}
  h=Encoder(x)
  \label{eq1}
\end{displaymath}
Then a CTC model and an attention based decoder generates objectives simultaneously based on the high-level feature $h$. In our experiments, the RNN Encoder is implemented by stacking multiple Bi-directional Long Short-Term Memory (BLSTM) layers. A detailed description of CTC and attention based decoder will be presented in Section 2.1 and Section 2.2 respectively. Then the hybrid CTC-Attention objective will be given in Section 2.3.

\subsection{Connectionist Temporal Classification (CTC)}
CTC provides a method to train RNNs without any prior alignment between input and output sequences of different lengths.
CTC introduces a latent variable, CTC path \(\pi=(\pi_1, \pi_2,\ldots, \pi_L)\), as the frame-level label of the input sequence. A special ``blank'' symbol is used to separate adjacent identical labels and represents a null emission.
By removing repetitions of identical labels and blank symbols, different paths can be mapped to a particular label sequence.
Based on the RNN encoder output $h$, CTC calculates the conditional probability of the label for each frame and assumes that the labels at different frames are conditionally independent.
So the probability of a CTC path can be computed as follows:
\begin{equation*}
  p(\pi|x)=\prod_{l=1}^{L} q_l^{\pi_l}
  \label{eq2}
\end{equation*}
where \(q_l^{\pi_l}\) denotes the softmax probability of outputting label \(\pi_l\) at frame \(l\). $q_l = \left( q_l^1, \cdots\, q_l^{K+1} \right)$ is often called the softmax output.
The likelihood of the label sequence is the sum of probabilities of all compatible CTC paths:
\begin{equation*}
  p(y|x)=\sum_{\pi\in\phi(y)} p(\pi|x)
  \label{eq3}
\end{equation*}
where \(\phi(y)\) denotes the set of all the CTC paths which can be mapped to the label sequence \(y\).

A forward-backward algorithm can be employed to efficiently sum over all the possible paths.
The likelihood of \(y\) can then be computed with the forward variable \(\alpha_l^u\) and the backward variable \(\beta_l^u\) as follows:
\begin{equation*}
  p(y|x)=\sum_u \frac{\alpha_l^u \beta_l^u}{q_l^{\pi_l}}
  \label{eq4}
\end{equation*}
where \(u\) is the label index and \(l\) is the frame index.
The CTC loss is defined as the negative log likelihood of the output label sequence:
\begin{equation*}
  L_{CTC}=-\ln(p(y|x))
  \label{eq5}
\end{equation*}
By computing the derivate of the CTC loss with respect to the softmax output \(q_l\),  the parameters of the RNN Encoder can be trained with standard back-propagation.

\subsection{Attention based Decoder}
The attention based decoder is an RNN which converts the high-level features $h$ generated by the shared encoder into the output label sequence with the attention mechanism. The decoder calculates the likelihood of the label sequence, based on the conditional probability of the label \(y_u\) given the input feature \(h\) and the previous labels \(y_{1:u-1}\), using the chain rule:
\begin{equation*}
  p(y|x)=\prod_u p(y_u|h,y_{1:u-1})
  \label{eq6}
\end{equation*}
At each step \(u\), the decoder generates a context vector \(c_u\) based on all the input features \(h\) and attention weight \(a_{u,l}\):
\begin{equation*}
  c_u=\sum_l a_{u,l}h_l
  \label{eq7}
\end{equation*}
The attention weight \(a_u=(a_{u,1}, a_{u,2},\ldots, a_{u,L})\) is obtained from location based attention energies \(e_{u,l}\) as follows:
\begin{displaymath}
\begin{split}
	a_{u,l} &=softmax(e_{u,l})\\
  e_{u,l} &=\omega^T tanh(W s_{u-1} + V h_l + M f_{u,l} + b)\\
  f_u &= F * a_{u-1}
\end{split}
\end{displaymath}
where \(\omega,W,V,M,b\) are trainable parameters, \(s_{u-1}\) is the decoder's RNN state. \(*\) denotes the one-dimensional convolution along the frame axis, \(l\), with the convolution parameter $F$, to produce the features \(f_u=(f_{u,1}, f_{u,2},\ldots, f_{u,L})\).

With the context vector \(c_u\), we can predict the RNN hidden state \(s_{u}\) and the next output \(y_u\) as follows:
\begin{displaymath}
\begin{split}
  s_u &=LSTM(s_{u-1},y_{u-1},c_u) \\
  y_u &=FullyConnected(s_u,c_u)
\end{split}
\end{displaymath}
where the $LSTM$ function here is implemented as a uni-directional LSTM layer and the $FullyConnected$ function indicates a feed-forward fully-connected network.

In the attention based decoder module, a special start-of-sequence symbol \(\langle sos \rangle\) and end-of-sequence symbol \(\langle eos \rangle\) has been added to the output sequence. When \(\langle eos \rangle\) is emitted, the decoder stops the generation of new output labels.

Finally, the attention loss is defined as the negative log likelihood of the target sequence.

\subsection{Hybrid CTC-Attention Objective}

With the aim to take advantage of both models, the CTC loss and attention loss can be combined \cite{kim2017joint}. We show the overall architecture of the hybrid model in Figure \ref{fig:speech_production}.

Both CTC and attention based methods their own drawbacks.
CTC makes the conditional independent assumption between the labels, thus requiring a strong external language model to compensate for the long term dependency between the labels. The attention mechanism produces each output using a weighted sum over all the input without any constraint or guidance which can be provided by alignments. Thus it is usually difficult to train the attention based decoder.

Note that the forward-backward algorithm in CTC can learn a monotonic alignment between acoustic features and label sequences, which can help the encoder to converge more quickly. Moreover, the attention based decoder can learn the dependencies among the target sequence. Hence, combining CTC and attention loss not only can help the convergence of the attention based decoder, but also enable the hybrid model to utilize label dependencies.

The hybrid CTC-Attention objective is defined as a weighted sum of CTC loss and attention based loss:
\begin{equation*}
  L_{hybrid}=\lambda L_{CTC} + (1-\lambda) L_{Attention}
  \label{eq13}
\end{equation*}
where \(\lambda \in (0,1)\) is a tunable hyper-parameter.

\section{Using Subword Units}

Traditional phoneme-based speech recognition systems require an external pronunciation lexicon to link phonemes and words. Thus, out-of-vocabulary (OOV) words (such as name entities or rare words) cannot be recognized. In addition, the pronunciation lexicon also complicates the decoding procedure.

End-to-end speech recognition can directly map acoustic frames to characters, words or subwords.
For word-based end-to-end system, the drawback is that OOV words can not be recognized and a large lexicon is needed, which suffers from expensive computation cost due to a large softmax output.
For character-based end-to-end system, the drawback is that the decoder's computation cost is increased and it is difficult to learn word-level dependency in the target sequence.

Subwords are chosen as the model units in our speech recognition system. Subword units are obtained by the byte-pair encoding (BPE) algorithm \cite{sennrich2015neural}, which iteratively merges the most frequent pairs of units (initially all are characters) and adds it into the set of subword units.
We define the initial subword set as the character vocabulary (`A',`B',$\ldots$,`Z') plus an additional symbol `\_' indicating the end boundary of a word.
For example, if `AB' is the most frequent pair of units in the current set, then `A' and `B' are merged to produce a new unit `AB', which will be added into the subword set. The iteration is ended until a given number of merging operations is reached.

Since the BPE algorithm maintains all the characters in the subword set, rare words can be represented by subword units.
Once we obtain the subword set, we break the word-based training transcripts into subword sequences, by greedily segmenting the longest subwords from left to right in a sentence.

\begin{table}[]
\centering
\caption{An utterance in the training data is segmented into words, characters and subwords respectively. The special symbol `\_' denotes the word boundary so that the original word sequence can be restored from character and subword based sequence. }
\label{tab:segmented_sequence}
\begin{tabular}{cc}
\hline
Basic Unit & Segmented Sequence                                                                                                                              \\
\hline
word       & \tabincell{c}{that neither of them had crossed \\ the threshold since the dark day   }   \\
character  & \tabincell{c}{t h a t \_ n e i t h e r \_ o f \_ t h e m \_ h a d \_ \\ c r o s s e d \_ t h e \_ t h r e s h o l d \_ s i n c e \_ \\ t h e \_ d a r k \_ d a y \_ }\\
subword   & \tabincell{c}{that\_ ne i ther\_ of\_ them\_ had\_ cro s sed\_ \\the\_ th re sh old\_ sin ce\_ the\_ d ar k\_ day\_}    \\
\hline
\end{tabular}
\end{table}

Table \ref{tab:segmented_sequence} shows an example of three different segmentation of an utterance from the training transcripts.
The original transcript is the word based sequence. We obtain the character sequence by simply dividing words into characters one by one and and adding the word boundary symbol `\_'.
It can be easily seen from Table \ref{tab:segmented_sequence} that the character based sequence representation is long, which is not good for decoding.
Using a greedy searching algorithm, the word sequence can be segmented into subword units, out of the 500 subword units generated by BPE.
We can see that frequent words, such as `that' and `of', and single character like `s' and `d', occur in the subword sequence. The subword based segmentation keeps the representation flexibility as the character based segmentation, but has a much shorter sequence length, which is good for decoding.

When encountering an OOV word such as ``cyberlife'', phoneme and word based systems mark this word with a special $\langle unk \rangle$ symbol, making recognition errors.
In character based systems, the decoder attempts to output character sequence (`c', `y', $\cdots$, `e', `\_') one by one. A substitution error would occur easily if any of these characters was wrong.
In subword based systems, since the subword `cyber' and `life' are frequent words or word roots, the decoder can predict the subword decomposition (`cyber', `life\_') with only two decoding steps and can recognize the rare word `cyberlife' more easily.

The subword sequence is used as the target output to train our hybrid CTC-Attention system. In decoding, a subword sequence is first produced by the decoder and then converted to the corresponding word sequence by removing the word boundary symbol `\_'.

\begin{table}[]
\centering
\caption{Word Error Rates (WERs) on the LibriSpeech subsets test\_clean and test\_other. The hybrid CTC-Attention (CTC+Att) model outperforms the pure CTC and Attention (Att) based model, when using characters (char). The combined parameter $\lambda$ is set to 1.0 for pure CTC and 0.0 for pure attention based experiments.
We extract 500 and 1000 subword units with the BPE algorithm. The hybrid CTC+Att model with 500 and 1000 subword units achieve the WERs 6.8\% and 7.6\% on the test\_clean set respectively, representing 12.8\% and 2.6\% relative improvements over the char baseline.
Note that no language models are applied in our experiments.}
\label{tab:main_results}
\footnotesize
\begin{tabular}{ccccc}
\hline
Model   & output unit   &   $\lambda$   & \multicolumn{2}{c}{WER}    \\
\hline
        &               &               & test\_clean  & test\_other \\
\hline
CTC     & char          &     1.0       & 20.9         & 39.8        \\
Att     & char          &     0.0       & 10.5         & 30.9        \\
CTC+Att & char          &     0.2       & 7.8          & 21.9        \\
CTC+Att & subword 500  &     0.2       & \textbf{6.8} & 19.5        \\
CTC+Att & subword 500  &     0.5       & 7.6          & 21.0        \\
CTC+Att & subword 1000 &     0.2       & 7.6          & 21.2        \\
\hline
 \tabincell{c}{wav2letter \cite{collobert2016wav2letter}\\4-gram LM} & char & - &7.2 & - \\
\hline
\end{tabular}
\end{table}

\section{Experiments}
\subsection{Experimental Setup}
The hybrid CTC-Attention based systems are experimented with the Chainer \cite{tokui2015chainer} backend of the ESPNET toolkit \cite{watanabe2018espnet}, using characters or subwords. CTC and attention-based systems are implemented by setting $\lambda$ to $1$ and $0$, respectively.
For comparison, no lexicon and language model are used in all the recognition systems.

We train and test different systems over LibriSpeech dataset \cite{panayotov2015librispeech}, consisting of 1000 hours of read audio books. The dev and test subsets of LibriSpeech are classified into two categories: simple (`clean') and hard (`other') subsets. We monitor convergence with LibriSpeech subsets dev\_clean and dev\_other. For evaluation, we report the word error rates (WERs) on the subsets test\_clean, test\_other. The acoustic features are 40 dimensional filterbanks generated by Kaldi \cite{povey2011kaldi}, with mean subtraction and variance normalization on a per-speaker basis.

The RNN encoder is a 8-layer BLSTM with 320 LSTM cell units per-direction, and each BLSTM layer is followed by a Batch Normalization layer \cite{ioffe2015batch}.
Each of the top two layers subsamples the hidden state with a factor of 2 from the output of the layer below.
The attention decoder is a 1 layer uni-directional LSTM with 320 units. 10 convolution filters of width 100 are used to compute location based attention energies. The Adadelta algorithm with gradient clipping is adopted as our optimizer, with hyper-parameter \(\epsilon=10^{-8}\).
For decoding, we use the beam search algorithm with the beam size 20.
All the experiments are performed with 4 Tesla K80 GPU. It takes about 2 days to train the hybrid model over the 1000h dataset.
Subword units are extracted using all the transcripts of training data by BPE algorithm. The number of subword units is set to 500.

\subsection{Results and Discussions}

Results of various systems are shown in Table \ref{tab:main_results}, from which we have the following main comments.

First, for all the different systems using characters, the hybrid system outperforms both the CTC and the attention based systems greatly, since it benefits from both loss.

Second, when using 500 subword units extracted from the training transcripts, the subword-based hybrid system obtains 6.8\% WER on test\_clean and 19.5\% on test\_other.
This represents a significant improvement (12.8\% and 7\% respectively) over the character-based hybrid system.

Third, we examine the effects of different $\lambda$.
The best tuned $\lambda$ is $0.2$, the same as in \cite{kim2017joint}.
Using a larger $\lambda=0.5$, the CTC loss forms a greater proportion in the hybrid loss. The WER degrades because CTC performs badly without an external language model.
Also note that the pure attention based system produces inferior performance without an auxiliary CTC Loss.

Forth, we examine the effects of using different number of subword units.
The performance of the hybrid system using 1000 subword units is slightly better than the character based hybrid system.
When increasing the number of subword units, the occurrences of subword units will become sparser.
For example, the least frequent unit `q' occurred 97 times in the 500 subword set; for the 1000 subword set, the least frequent unit becomes `toge' and occurs only 3 times in the training data.
As a result, the model performance deteriorates due to the data sparseness problem.
A larger number of subwords, such as 10k, is more suitable for tasks with larger training corpus, like machine translation \cite{wu2016google}.

Finally, we examine the performances of the subword based hybrid system, using different amount of training data.
Two subsets of training data are drawn randomly from the 1000h Librispeech training set, having 100h and 500h respectively.
We use the experimental setup which corresponds to the subword set size of 500 and $\lambda$ of 0.2 in Table \ref{tab:main_results}.
The results are shown in Table \ref{tab:hour_diff}.
The poor results from using small-sized training dataset indicate that we need a large-sized training dataset in order to successfully train subword systems.
It can be seen that the WER decreases rapidly when increasing the size of the training data.
Thus, the subword systems potentially can perform even better when trained over much larger scaled data.

\begin{table}[]
\centering
\caption{Experiments with different sizes of training data. We randomly select 100h and 500h from the 1000h LibriSpeech dataset. All experiments use the same setup which corresponds to the subword set size of 500 and $\lambda$ of 0.2 in Table \ref{tab:main_results}.}
\label{tab:hour_diff}
\begin{tabular}{ccccc}
\hline
hours          & \multicolumn{2}{c}{WER}    \\
\hline
               & test\_clean  & test\_other \\
\hline
100            & 34.7            & 45.4        \\
500            & 10.4            & 26.9        \\
1000           & 6.8          & 19.5        \\
\hline
\end{tabular}
\end{table}

\section{Conclusions and Future Work}
In this work, we present a hybrid CTC-Attention based end-to-end speech recognition system using subword units, which works without any dictionary and language model.
Compared to the character-based hybrid system, the proposed subword-based hybrid system significantly reduces WERs in both clean and noisy conditions.

Given the demonstrated benefits of using subword units, it is worthwhile to further study techniques for better subword unit construction and decompositions. Another future work is to apply the proposed system to larger scale speech recognition tasks with tens of thousands hours of speech data.


\bibliographystyle{IEEEtran}

\bibliography{mybib}

\end{document}